\documentclass[12pt]{article}
\begin{document}

\title{{\bf Scientific and Philosophical Challenges to Theism}
\thanks{Alberta-Thy-21-07, arXiv:0801.0247, to be published in Melville
Y.~Stewart, ed., {\em Science and Religion in Dialogue} (Blackwell
Publishing Inc., Oxford), and in Melville Y.~Stewart and Fu Youde, eds.,
{\em Science and Religion: Current Dialogue} (Peking University Press,
Beijing, in Chinese), from a series of lectures sponsored by the
Templeton Foundation and given at Shandong University in Jinan, China,
autumn 2007; see also arXiv:0801.0245 and arXiv:0801.0246.}}

\author{
Don N. Page
\thanks{Internet address:
don@phys.ualberta.ca}
\\
Institute for Theoretical Physics\\
Department of Physics, University of Alberta\\
Room 238 CEB, 11322 -- 89 Avenue\\
Edmonton, Alberta, Canada T6G 2G7
}
\date{(2008 February 14)}

\maketitle
\large
\begin{abstract}
\baselineskip 20 pt

\hspace{.20in} Modern science developed within a culture of
Judeo-Christian theism, and science and theism have generally supported
each other.  However, there are certainly areas in both science and
religion that puzzle me.  Here I outline some puzzles that have arisen
for me concerning everlasting life, human free will, divine free will,
the simplicity and probability of God, the problem of evil, and the
converse problem of elegance.

\end{abstract}
\normalsize

\baselineskip 19.4 pt

\newpage

\section{Introduction}

\hspace{.20in} Modern science developed within a culture of
Judeo-Christian theism for several reasons \cite{Hooykaas}.  For example,
the idea of a lawgiver for nature (i.e., God) encouraged belief in laws
of nature.  Also, the need to study the laws of nature was encouraged by
the Biblical command in the first book of the Bible, Genesis 1:28:  ``Be
fruitful and multiply; fill the earth and subdue it; have dominion over
the fish of the sea, over the birds of the air, and over every living
thing that moves on the earth'' \cite{Bible}.

But once the idea of laws of nature was derived from the idea of a
lawgiver, one could often forget the lawgiver and just study the laws,
rather as citizens in a nation can obey its laws without thinking about
who made those laws.  There is then a tendency to conclude that there is
no lawgiver at all.

Since both science and religion are human activities, and since humans
often have conflicts, it is not surprising that there are
science-religion conflicts.   Since science and religion tend to claim
jurisdiction over territories that have historically overlapped, it is no
wonder that conflict should on occasion have arisen between them, as the
aftermath of the Galileo affair would illustrate \cite{Galileo}.

At other times there are genuine human uncertainties and differences of
opinion.  For example, theists have differed over whether the evidence
for biological evolution is convincing, though now it seems that most
theologians accept it.  Somewhat similarly, today there is disagreement
within both theists and scientists about whether the multiverse ideas are
correct \cite{multiverse}.

Generally I see science and religion as supporting each other, but there
are certainly areas in both that puzzle me.  Let me discuss some that to
me have seemed to be the biggest challenges to theism, and give some
thoughts I have had on them.  These thoughts are certainly tentative, so
I would certainly appreciate any help others can provide on these
mysteries.

\section{The Afterlife Awareness Problem}

\hspace{.20in} One rather arcane challenge that has occurred to me is the
application of the Carter-Leslie-Nielsen-Gott doomsday argument
\cite{C,L2,N,L3,G,L16} to the afterlife.  The original doomsday argument
is that the observation that we are among the first hundred billion or so
humans reduces the prior probability that we find ourselves in a species
whose total lifetime number of individuals is much higher.  If humans
were to continue at present or growing populations for more than a few
hundred additional years, it would be unlikely for us to have found
ourselves in the very small fraction alive by now.  On the other hand, if
the human race were to end sooner, we would not be so unusual.

The doomsday argument implies that unless the {\it a priori} probability
is very high for far more humans in the future than in the past, then our
observations of how many humans there have been in the past makes the
{\it a posteriori} probability low for far more humans in the future. 
Although this argument has been widely debated and disputed (too widely
for me to give a comprehensive list of references), it has never been
refuted, and I believe that it is basically valid.

I realized several years ago that similar consequences would seem to
apply to hypotheses about an afterlife, experiences after physical
death.  If one were not absolutely certain of an afterlife that would
last enormously longer than the pre-death life, then the observation that
we are experiencing (presumably) pre-death life rather than the afterlife
would significantly reduce the {\it a posteriori} epistemic probability
for a very long afterlife.  Otherwise, our present experiences would seem
to be highly unusual if there were in fact far more afterlife experiences
than pre-death experiences.

After puzzling over this for several years (and having long series of
email discussions with a small number of people, most particularly
Richard Swinburne, who were kind enough to consider my thoughts on it
without necessarily agreeing with the presuppositions), I stumbled upon
an analogous possibility in physics, the formation of brains by vacuum
fluctuations, which gives a similar problem if the universe lasts too
long \cite{bb1,bb2,bb3}.  These brains are rather similar to those which
had been earlier proposed to arise from thermal fluctuations and which
had been named Boltzmann brains \cite{BB}, following a somewhat analogous
suggestion by Boltzmann \cite{Boltzmann} that he actually attributed to
his assistant Dr. Schuetz (leading Andreas Albrecht, the originator of
the phrase Boltzmann Brain, to quip that they might actually be better
named Schuetz's Schmartz).  Therefore, I generally followed the usage of
this catchy name for my related idea of brains as vacuum fluctuations,
though so far as I know I was to first to raise the problem with vacuum
fluctuations themselves.

The problem with these generalized Boltzmann brains is that if the
universe lasts too long (e.g., infinitely long), then per comoving volume
(i.e., in a region that expands with the universe and would contain a
fixed number of atoms if they are not created or destroyed), there would
only be a fixed number of ordinary observers (conscious beings like us
who presumably evolved by natural selection), since they can surely last
only a finite time, say when there are still stars burning, but on the
other hand there would be a much larger (e.g., infinite) number of
Boltzmann brains.  Then almost all observations per comoving volume would
be made by Boltzmann brains.  So if this scenario were correct, we would
most likely be Boltzmann brains.

But Boltzmann brains are very unlikely to observe the order that we
actually see, so our ordered observations (with rather coherent detailed
memories, etc.) are strong evidence against our being Boltzmann brains
and therefore also against there being far more Boltzmann brains than
ordinary observers.  In this way we have observational evidence against a
universe that lasts too long, at least under the assumption that
Boltzmann brains can form from vacuum fluctuations at similar rates at
any time arbitrarily far into the future, and under the assumption that
we count the ratio of Boltzmann brains to ordinary observers in a fixed
comoving volume before our universe possibly tunnels into something
different.

There are many conceivable solutions to the Boltzmann brain problem
\cite{BBsol1,BBsol2,BBsol3,BBsol4,BBsol5,BBsol6,BBsol7,BBsol8,BBsol9,
BBsol10,BBsol11,BBsol12,BBsol13,BBsol14,BBsol15,BBsol16,BBsol17,BBsol18,
BBsol19,BBsol20,BBsol21,BBsol22,BBsol23,BBsol24,BBsol25,BBsol26}, some of
which abandon the assumptions of the previous paragraph, though none that
is universally accepted.  For the analogous problem with the afterlife,
which I might call the afterlife awareness (AA) problem instead of the
Boltzmann brain (BB) problem, what seems to me possibly most relevant is
the solution I suggested \cite{bb1,bb2,bb3} that instead of persisting
indefinitely into the future, our universe could be decaying quantum
mechanically at a rate comparable to its exponential expansion rate,
decaying at least fast enough to make the expectation value of the
four-volume of the comoving region (the three-volume multiplied by the
persistence probability and integrated over the time) finite and not so
large that Boltzmann brains would dominate over ordinary observers
(OOs).  This proposed solution has its own problem of the fine tuning of
the decay rate (which is not one of the fine tunings that might be
explained by the observational selection of the anthropic principle),
since the decay rate has to be great enough and yet not so great to make
it highly improbable that the universe has lasted as long as it already
has.

However, whether or not my suggested solution to the Boltzmann brain
problem in our physical universe is correct, it occurred to me that
something similar might conceivably be a solution to the analogous
afterlife awareness problem.  But before explaining this, I need to
reformulate the problem in terms of the measures of conscious experiences.

All of the problems discussed here, the doomsday (DD) argument, the
Boltzmann brain (BB) problem, and the afterlife awareness (AA) problem,
arise from comparing the probability of an experience to be one of those
(which does not fit our observations) to that of being an ordinary
observer (OO) (which apparently would fit our observations).  (I am
tempted to think of other related problems, such as that of CCs or
conscious clouds \cite{Hoyle}, but I haven't yet thought of analogous
problems for all of the letters of the alphabet after D to O, though it
did occur to me that EEs could be eternal experiences, conscious
experiences that are each eternal, unlike our experiences that are each
momentary.)

One might well postulate that each conscious perception or experience
(all that one is momentarily conscious aware of) has a {\it measure}
associated with it, giving the probability of that experience's being
selected at random if a random selection were made
\cite{SQM1,SQM2,Bostrom,MS}.  (One does not need to suppose that there
actually {\it is} any such random selection in order to be able to
calculate likelihoods {\it as if} such a selection occurred.)  Then any
set of conscious perception would have a corresponding measure obtained
by summing the measures for the individual perceptions.  (It could be the
case that conscious perceptions form a continuum rather than a discrete
set, in which case one would only have positive measures for continuous
sets of perceptions, rather than for individual perceptions, analogous to
the way that there is positive volume only for continuous sets of spatial
points and not for the individual points.  But for simplicity here I
shall consider just the alternative logical possibility that the
perceptions are discrete rather than continuous.)

In a classical physics picture of a single conscious being having a
unique temporal sequence of perceptions, one might make an approximation
that the measure for a particular sequential set of perceptions is
proportional to the time taken to have that set.  So if during 16 waking
hours (960 minutes) in a day one is conscious of eating for 96 minutes,
say, one might take the measure for experiences of eating during that day
to be 0.1 or one-tenth of the measure for all the conscious experiences
of the waking hours.  (I don't mean to be denying that one is having
conscious perceptions while asleep during dreams, but I am avoiding the
question of how to compare their measure with perceptions while awake.)

In a quantum physics picture, there are also what are usually interpreted
as the quantum probabilities of various alternative possibilities.  So,
for example, if there is a quantum probability of 0.2 that one is fasting
during those 16 waking hours and 0.8 that one spends 96 minutes eating,
then the relative measure of the experiences of eating would be only 0.8
times 0.1, or 0.08.  Or, if one would be eating caviar for 96 minutes in
a day if one won a lottery for which the quantum probability of winning
is one in a million (and if one would not be eating caviar at all
otherwise), the relative measure of the experience of eating caviar, out
of all the experiences in that day with all results of the lottery, would
be $10^{-7}$.

I have formulated a framework for connecting consciousness to physics
\cite{SQM1,SQM2,MS} in which each conscious perception has a measure
given by the expectation value of a corresponding quantum operator. 
However, for the present discussion, it is not important whether or not
this framework is correct, but just that conscious perceptions do have
objective measures that give frequency-type probabilities of the
perceptions' being selected if they were selected at random.  (Since I
believe that the random selection is purely hypothetical, these
probabilities, while being perfectly objective for each possible theory
giving them, are also hypothetical, what I might call objective
hypothetical probabilities, to be distinguished from the subjective
epistemic probabilities that one might assign to various theories that
are not known to be correct or incorrect.)

Now suppose that one has various theories $T_i$ that each tell what
fraction $f_i$ of the measure of all conscious perceptions or experiences
are pre-death rather than afterlife.  (For simplicity I shall just focus
on these two possible types and also assume that the content of each
experience clearly identifies which type it is, leaving out such
experiences as those here in the present life in which one might feel
that one has gone to heaven or hell.)  Then in each $T_i$, the
probability that a randomly selected conscious experience would be a
pre-death experience would be $f_i$.  Given the information that one is
observing a pre-death experience (and no other information), this $f_i$
would then be the likelihood of the theory $T_i$.  (If one includes other
information from the particular conscious perception being experienced,
that would further restrict the fraction and give a lower likelihood.)

For a theory $T_i$ predicting no afterlife, all experiences would be
pre-death, so $f_i = 1$.  For a theory predicting an afterlife with far
greater measure of experiences, $f_i \ll 1$.  In the limit of an infinite
measure of experience for afterlife awarenesses but still only a finite
measure for pre-death experiences, $f_i = 0$.

Now suppose that various theories of these different types are all
assigned nonzero {\it a priori} probabilities, so that one is originally
not epistemically certain of any of them (though one might have strong
preferences).  Then by Bayes' theorem, the final epistemic {\it a
posteriori} probabilities to be assigned to the theories would be
proportional to their {\it a priori} probabilities multiplied by their
likelihoods $f_i$.  This then has the effect of reducing the {\it a
posteriori} probabilities more for the theories in which the fraction of
the measure of pre-death experiences is smaller.  For a theory in which
$f_i$ is very tiny (a very large relative measure of afterlife
awarenesses), the {\it a posteriori} probability would be less than 1/2
unless the total {\it a priori} probability of all the theories with
higher $f_i$ is smaller than $f_i$.  In particular, if the total {\it a
priori} epistemic probability of all the theories with $f_i = 1$
(theories with no afterlife; all experiences pre-death experiences) is
positive, larger than zero, then in the limit that any theory gives $f_i
= 0$ (i.e., by having an infinite measure of afterlife awarenesses and
only a finite measure of pre-death experiences), it would have zero {\it
a posteriori} epistemic probability.

Thus it seems that unless one started absolutely certain of an infinite
afterlife, after considering the evidence that one is having a pre-death
experience instead, one should then assign zero epistemic probability to
the idea of an infinite afterlife.

Given that I had previously had faith in an infinite afterlife, though
not quite 100\% faith, this conclusion certainly seemed contrary to how I
had interpreted the afterlife.  It has bothered me ever since I first
thought of it.

One possible solution, suggested to me by the originator of the doomsday
argument, Brandon Carter \cite{Carterpriv}, is that the afterlife is not
an infinite set of afterlife awarenesses (AAs) but a single eternal
experience (EE), or a single one for each person.  This finite number of
eternal experiences could then have a large but finite measure, leaving
$f_i$ and the resulting {\it a posteriori} probability nonzero (assuming
one was not absolutely certain that this theory is wrong and so assign it
zero {\it a priori} probability).  Perhaps this would be a more
sophisticated way of looking at eternal life, not as an infinite set of
experiences but as one single eternal experience, or as one single
eternal experience for each person.

Another suggested solution was given me by the person who has expounded
the doomsday argument the most thoroughly, John Leslie \cite{Lesliepriv},
following arguments by Andrei Linde that Leslie previously disputed
\cite{L16}.  Linde's argument applied to an infinite afterlife would be
that although any pre-death experience would be infinitely early if there
is an infinite afterlife, it would not be {\it specially} early, since
all experiences at finite time, even in the afterlife, would also be
infinitely early.  Maybe Leslie's change of mind in now accepting this
argument is correct, but to me the pre-death experiences would still seem
to have zero probability in comparison with an infinite measure of
afterlife awarenesses, so it does not solve the problem in my own mind.

Another point made by Leslie \cite{Lesliepriv} is that it is just
subjective epistemic probabilities that are being discussed, not true
objective probabilities that are ``out there'' in reality.  That is, it
presumably is the case that either an afterlife of infinite measure
definitely exists or definitely does not exist, so the objective
probability is either 1 or 0.  As Leslie notes, ``God's infinitely
powerful and benevolent!''  Therefore, one might remain confident that
the probability of an infinite afterlife is really 1, even though
applying a Bayesian analysis to an uncertain epistemic {\it a priori}
probability (not quite unity for an infinite afterlife) might give zero
{\it a posteriori} epistemic probability for it.  Nevertheless, it does
bother me that a line of Bayesian reasoning that apparently works in
finite cases seems to give zero {\it a posteriori} epistemic probability
for a theistic doctrine that I formerly was much more certain about.

However, another possible way to solve the afterlife awareness problem
and avoid the infinite measure of those experiences that would give $f_i
= 0$ for the presumably finite measure of pre-death experiences (or at
least finite measure of pre-death experiences per person), would be to
have the measures of each of the infinitely many AAs not constant but
decaying sufficiently rapidly in some ordering of them.  That is, if one
orders the afterlife in decreasing order of their individual measure, one
could have the measure per AA decreasing fast enough that the sum
converges to a finite total measure that is not too much greater than the
corresponding sum of the measures of all the pre-death experiences.

For example, one might postulate that in some theory $T$, there is a
countably infinite set of AAs, and the $n$th AA has the measure
$A(1-x)x^{-n}$ for some $x$ between 0 and 1, so that the sum over $n$ is
the finite total measure $A$ of the AAs.  Then if the sum of the measures
of the pre-death experiences in this theory is $B$, the fraction of the
measure that is pre-death is $f = B/(A+B)$.  This is the likelihood of
this theory under the observation of a pre-death experience.

Assume for simplicity that this is the only theory under consideration in
which there is an afterlife, and that its {\it a priori} probability is
assigned to be $p$.  Then all the other (non-afterlife) theories have
total {\it a priori} probability $1-p$ and have unit likelihood under the
observation of a pre-death experience (since all of their experiences are
pre-death).  The product of the {\it a priori} probability and the
likelihood for this afterlife theory $T$ is then $pf$, and the sum of the
products of the {\it a priori} probabilities and the (unit) likelihoods
for all the non-afterlife theories is $1-p$.  Normalizing by the sum of
these products, which is $pf+1-p$, Bayes' theorem gives the {\it a
posteriori} probability of this afterlife theory $T$ as $pf/(pf+1-p) =
pB/(A+B-pA)$, which is not too small so long as $pf$ is not too small in
comparison with $1-p$, the total {\it a priori} probability of all the
(non-afterlife) theories.  In particular, the afterlife theory $T$ has an
{\it a posteriori} probability greater than 1/2 if its {\it a priori}
probability is $p > 1/(1+f) = (A+B)/(A+2B)$, or if $A/B < (2p-1)/(1-p)$. 

If one is nearly certain {\it a priori} of this afterlife theory $T$, so
$p$ is near unity, then the afterlife can have much more measure $A$ than
the pre-death measure $B$ and yet still give an {\it a posteriori}
probability greater than 1/2.  For example, if one were initially 99\%
certain of the afterlife, $p=0.99$, then its {\it a posteriori}
probability would be greater than 1/2 for all $A$ up to $98B$, a total
afterlife measure up to 98 times that of the pre-death measure.  However,
if one initially had only 60\% confidence in the afterlife, then its {\it
a posteriori} probability would be greater than 1/2 only for all $A$ up
to $B/2$, a total afterlife measure only up to half that of the pre-death
measure.

If such an afterlife theory were true, one could have an infinite number
of afterlife experiences and yet their total measure would not
necessarily swamp that of the pre-death experiences to such an extent
that it would have low {\it a posteriori} probability in view of our
observation of having a pre-death experience.  Of course, the total
measure would not be infinitely greater than the pre-death experiences,
and the probability for finding oneself experiencing an AA far along the
sequence of ever-decreasing measure would be very small.  

In this proposed possible solution of the AA problem, earthly pre-death
experiences would not be an infinitesimal fraction of the total, though
they could still be a rather small fraction of the total.  Furthermore,
the quality of the afterlife experiences could presumably be arbitrarily
more intense than those of our present pre-death experiences.  Although
the New Testament and the Koran stress the superiority of heaven, I do
not see that they say it is infinitely more important than life, justice,
and righteousness here on this earth.  Therefore, it is not obvious that
this solution to the AA problem is incorrect, though it is certainly
speculative and highly tentative, since to the best of my knowledge both
the AA problem and its possible solutions have not been discussed in the
traditional theological literature.

\section{Human Free Will}

\hspace{.20in} Another potential problem, or at least controversial issue
within theistic beliefs, is the question of human free will, which has
often been invoked to explain human responsibility and the existence of
evil caused by humans.  Is there any room for human free will in a
universe with definite laws of nature and a definite quantum state? 
I.e., if the initial conditions and the dynamical laws of evolution are
determined, how could humans act otherwise than what would be predicted
by these initial conditions and dynamical laws?  (Here I am taking free
will in the libertarian or incompatibilist sense of being incompatible
with determinism or complete determination ultimately by causes or
entities other than the being to whom the free will is ascribed.)

One logically possible answer is that human free will could help choose
the laws and the quantum state of the universe.  But since that would
involve determining the quantum state of the very early universe, far
before humans existed, that would seem rather implausible.  If one
assumed that causality acts only forward in time, and if time indeed goes
forward from the early universe to the existence of humans, this logical
possibility would seem to be physically impossible.  However, we do not
fully understand the nature of causation, so it is not completely obvious
that causality backward in time really is physically impossible.  Indeed,
the ordinary concept of causality in physics says that the state of a
closed system is completely determined by the laws of evolution and the
state at any time, so from the state at one time, the state at all times,
both before and after, would be determined, and therefore there is no
obvious restriction just to causality forward in time.  Nevertheless,
even though I do accept timeless views of the universe in which there is
no fundamental asymmetry in time and no time asymmetry in causality, I do
personally find it rather implausible that human free will choices can
help determine the quantum state of the universe from the very beginning.

However, I find an even stronger argument against human free will to be
the realization that if God creates everything logically contingent other
than Himself, then free will by any created being seems to me to be
logically impossible.  That is, I see the concept of free will by created
beings to be a logical contradiction to the concept that they are created
{\it ex nihilo} by God (or even just to the concept that one can trace
back all causation to God).  For if God entirely creates or ultimately
causes everything contingent other than Himself, as I believe, He creates
or causes all such entities not just as one time but at all times,
including all actions of the beings He creates.  If God creates or causes
everything other than Himself that does not exist necessarily, then it
seems to me that nothing other than necessary entities (e.g., logical
tautologies) can lie outside the purview of this creation or causation,
including any actions or choices by created beings.  If God totally
creates us or causes our entire existence, that would seem to imply that
He creates or causes everything that we do, so that all we are and do
would be completely created or caused, and hence determined, by God, with
our having no true (libertarian) free will.

Now I will admit that if we had some independent existence and were not
entirely created or caused by God, then logically we could have free
will.  God might adopt us, or at least our independent free will choices,
within a universe that He otherwise creates.  (I don't even see a
contradiction between this logical possibility and God's always knowing
what free will choices we would make, since the only contradiction I see
is with His creating or ultimately causing all that we are and do and the
claim that what we do is free from His determination.)  However, this
adoption picture, that our free will choices have some existence
independent of God and were adopted by God within His universe, seems to
leave God's not creating or causing everything contingent other than
Himself and hence seems less simple than the traditional monotheistic
view that God creates everything contingent other than Himself.

\section{Divine Free Will and Information Content}

\hspace{.20in} So far I have implicitly assumed that God Himself truly
has libertarian free will and can do whatever is not logically
inconsistent (though I am arguing that it seems logically inconsistent
for Him to create beings with libertarian free will, and, if so, that is
truly impossible for Him).  Sometimes it is assumed that if the laws of
nature are fixed and if the initial conditions or quantum state are also
fixed, then even God would not be free to make things otherwise.  For
example, when the Hartle-Hawking `no-boundary' proposal for the quantum
state of the cosmos \cite{Hawking,HH} was first proposed, I was defending
it at a small gathering of quantum cosmologists \cite{Nature}, and the
late Bryce DeWitt, often considered the father of quantum cosmology,
objected, ``You don't want to give God any freedom at all!''  But before
I could think of an answer, Karel Kucha\v{r} responded, ``But that's His
choice.''  In other words, even if we correctly deduced the quantum state
of the universe, it would have been God's choice to create the universe
in that state.  I.e., God's determining the universe to be in a
particular deterministic state would not contradict His free will in
making this determination, though it would seemingly preclude the
independent free will of any creatures created by God in this state.

However, there is a strand of traditional monotheistic thought, going
back at least to Anselm \cite{Anselm}, that God Himself is a necessary
being.  If that is interpreted to mean that God is an entirely necessary
being, than even God has no free will.  Furthermore, if the necessity of
God includes all of His activity, such as His activity in creating the
universe or multiverse, then the created universe or multiverse is also
necessary and not contingent.  Assuming that God indeed creates
everything otherwise contingent other than Himself (e.g., leaving out the
apparent logical possibility of other partially independent beings with
independent free will, perhaps themselves contingent rather than
necessary, that God might adopt within His creation without directly
creating their free will choices), then the entirety of existence, what
philosophers call the {\it world}, would be necessary \cite{Strand-Kraay}.

However, it is not clear to me that God must be a completely necessary
being.  Anselm's ontological argument just seems to imply the necessary
existence of the greatest necessary being, but if only tautologies such
as mathematical theorems have necessary existence, then Anselm's proof
would only imply that the greatest tautology necessarily exists, and not
what one would traditionally interpret to be God.  Therefore, it seems to
me that, so far as we know, God and the universe might be at least
somewhat contingent, not necessarily the way they actually are.

We might consider these various conceptual possibilities in terms of the
information content of God.  (I say ``conceptual possibilities'' to
denote concepts that we are not sure are impossible, since if in fact God
is a necessary being, then it is necessarily impossible for Him not to
exist, and on the other hand, if in fact God is not a necessary being,
then it is necessarily impossible for Him to have necessary existence. 
So one or the other of these ``possibilities'' presumably must be the
case, and the other necessarily impossible rather than really being a
``possibility,'' but it is just that we don't know for certain which is
necessary and so may epistemically regard both as ``conceptual
possibilities.''  I suppose an atheist might also raise the conceptual
possibilities not only that God might contingently not exist but also
that God might necessarily not exist.  I recall hearing this question
being asked, perhaps without realizing the full philosophical content, by
Lucy Hawking when she was quite young:  ``Is God impossible?'')

If God were entirely necessary, He would have no information content,
since the information content of an entity is the minimum that needs to
be specified in order from that information to deduce fully the
properties of the entity.  On the other hand, if God were contingent but
simple, He would have small but nonzero information content.  And yet a
third conceptual possibility is that God is contingent and irreducibly
complex, having a large information content.

\section{The Complexity and Probability of God}

\hspace{.20in} This third conceptual possibility is what Richard Dawkins
assumes in his popular book, {\em The God Delusion} \cite{Dawkins}.  For
much of the book, Dawkins sounds like an Old Testament prophet railing
against idolatry, except that he believes the worship of any God or gods
is idolatrous.  But the philosophical heart of his argument is Chapter 4,
where he argues against the existence of God by saying that God would
have to be extremely complex.  Since his arguments are not very tightly
stated, I formulated the heart of Dawkins' argument as a syllogism and
then, with the help of an email exchange with several colleagues,
especially William Lane Craig, I revised it to the following form:

\begin{enumerate}
\item A more complex world is less probable than a simpler world.
\item A world with God is more complex than a world without God.
\item Therefore a world with God is less probable than a world without
God.
\end{enumerate}

After circulating this form, I did get the obviously hurried reply from
Dawkins:  ``Your three steps seem to me to be valid.  Richard Dawlkins
[sic]'' (1 February 2007).

Now that I have summarized Dawkins' basic argument in a brief form that
he seems to agree with, modulo typos, one can ask whether Dawkins is
right.  The conclusion of the syllogism seems to follow from the two
premises (or at least I have intended this to be the case), so it is a
question of whether the premises are correct.

One might question whether complexity is improbable, an unproved
assumption.  There is also the fact that complexity depends on background
knowledge and may be only subjective.  For example, David Deutsch
\cite{Deutsch1} has emphasized to me that ``complexity cannot possibly
have a meaning independent of the laws of physics. \ldots If God is the
author of the laws of physics (or of an overarching system under which
many sets of laws of physics are instantiated---it doesn't matter) then
it is exclusively God's decision how complex anything is, including
himself.  There neither the idea that the world is `more complex' if it
includes God, nor the idea that God might be the `simplest' omnipotent
being makes sense.''  

This argument makes sense to me, but it did have the effect of shaking my
fundamentalist physicist faith in the simplicity of the laws of nature. 
However, more recently Deutsch has pointed out \cite{Deutsch2} that these
considerations do not mean that the concept of the simplicity of the laws
of physics is circular:  ``I don't think it's circular, because the fact
that simplicity is determined by the laws of physics does not mean that
all possible laws are `simple' in their own terms.''  So I suppose one
might still ask whether in a universe apparently governed by simple laws
of physics, God appears to be simple.  However, since as Deutsch notes,
God could have made Himself appear to have arbitrary complexity, it is a
bit dubious to say that His probability is determined by His complexity.

Nevertheless, since we scientists (and indeed most others) prefer
hypotheses that are ultimately simple, we might for the sake of argument
grant the first premise I have ascribed to Dawkins and ask whether the
second premise is correct.  Again Deutsch's comments should lead us to be
cautious in drawing such conclusions.  However, even if we take the
na\"{\i}ve view that one can define the complexity of God (say with
respect to the laws of physics in our universe), then it is still not
obvious that God is complex, or that He would add complexity to the
world.  Perhaps God is indeed simple \cite{Swinburne}.

If God were necessary, then He would have no complexity at all.  Even if
God were contingent, He might be simple.  For example, perhaps God is the
best possible being (assuming sufficient background knowledge that this
apparently simple definition uniquely specifies some possible entity,
though it is certainly unclear that our background knowledge within this
universe is sufficient for this).  Even if it is not necessary for such a
God to exist, He might be simple (if simplicity can indeed be defined).

Even if one concedes that the philosophical idea of God might be simple,
there is the question of whether God is simple in traditional
monotheism.  At first sight, the God of the Bible and of the Koran seems
complex.  But analogously, Earth's biosphere seems complex.  However, the
full set of biospheres arising by evolution in a huge universe or
multiverse with simple laws of physics might be simple.  Similarly, the
limited aspects we experience of God might be complex, but the entirety
of God might be simple.

\section{The Problem of Evil and Elegance}

\hspace{.20in} Perhaps the most severe problem of traditional theism is
the problem of evil.  If God is the best possible being and created
everything, why does evil exist?

If instead of being totally created and determined by God, we were
adopted and bring in evil by our own free will choices, this might
explain human evil, but one would still have the problem of natural
evil:  disease, earthquakes, storms, floods, and other evils not caused
by humans.  So whereas human free will is often invoked to solve the
problem of evil, it does not seem to give a full solution, and therefore
I do not regard the problem of evil as sufficient for me to give up my
simple hypothesis that God created and determined everything contingent
other than Himself.

Perhaps because I independently stumbled upon it myself, though many
years later, I regard the best tentative solution for the problem of evil
to be the multiverse theodicy \cite{Adams,McHarry,Perkins,Forrest81,
Ward,Stewart86,Coughlan,Leslie89,Stewart93,Forrest96,Leslie01,Turner,
Draper,Hudson,Kraay} that God created all universes that are better to
exist than not to exist.  So rather than God's just creating one or more
universes that have no evil, one might imagine that God thought it better
to create all universes that are better to exist than not to exist.  In
other words, instead of minimizing evil by avoiding creating any
universes with evil, God might be seeking to maximize the net good over
evil.  Therefore, instead of leaving our universe uncreated because it
has evil in it as well as good, God might have seen that it has more good
than evil and decided that it would be better to exist than not to exist.

I would think that it is certainly common to make analogous judgments
about the existence of persons rather than of universes.  For example, I
have certainly done evil and hurt other people.  Yet I still feel that it
is better for me to exist than not to exist.  I believe that it is the
same for you and hope that you feel similarly, that it is indeed better
for you to exist than not to exist.  If an entity has good that is
accompanied by a lesser amount of evil, then it indeed seems better for
that entity to exist, rather than that all evil be eliminated.

For me, particularly as one who has worked with and lived with the
wheelchair-bound Stephen Hawking, one of the most horrific scenes in the
movie {\em The Pianist} was when the Nazis entered a Jewish apartment in
Warsaw and ordered the people sitting around the table to stand.  An
elderly man in a wheelchair could not comply, so the Nazis then heaved
him out the third floor window to his death on the street below.  In
realizing that the inhuman Nazis were not actually nonhuman but exhibited
some of the same sinful tendencies that I see in myself, it occurred to
me that perhaps in their twisted minds they were trying to eliminate what
they saw as evil, disabilities and weaknesses.  Without endorsing the
Nazis' ideas of what is good and evil, which were also quite distorted, I
would indeed be sympathetic to efforts to reduce disabilities and
weaknesses in a person, with his or her consent, while also working to
enhance the good that the person has.  However, the Nazis' procedure of
cruelly eliminating entire persons and communities that they saw as
having evils or weaknesses (or often just differences from themselves)
was barbaric and totally unjustified.  Perhaps analogously, we should not
expect God to eliminate (or avoid creating) entire universes that have
both good and evil in them.

Now although I do believe that the many-universes solution to the problem
of evil is the best one I have heard of, I also think that so far it has
not completely solved the problem.  With this solution, one might expect
that if our universe is a typical one with more good than evil, it would
not have enormously more good than evil.  That is certainly consistent
with my general impression as a human of the moral goodness and evil on
the earth.  However, if one applies the same idea to the elegance of the
laws of physics as another good that God might be seeking to promote, one
might expect that the laws of physics of our universe would have more
elegance than ugliness, but not enormously more.  On the other hand, my
impression as a scientist is that the laws of physics are enormously more
elegant than ugly, so it seems doubtful that God created all universes
with just more elegance than ugliness in the natural laws.

In other words, for me the problem of evil (which might be explained by
having God create all universes better to exist than not) has been
replaced by the problem of elegance (why on the level of the beauty of
the laws of nature our universe seems enormously more elegant than
ugly).  Another expression of the problem is the question of why God
seems to have so much higher standards of mathematical elegance (not
allowing our universe much mathematical ugliness) than of moral good and
evil (apparently permitting a much higher ratio of moral evil to good).

\section{Conclusions}

\hspace{.20in} Challenges to theism go back to one of the oldest books of
the Bible, the Book of Job, whose lead character wrestled with the
problem of evil that was basically left unexplained to him.  As finite
beings, like Job we should not expect to understand everything, though it
is good to seek as much understanding as possible.  We can wrestle with
the problems, but in the end we have to live life with the limited
knowledge that we do have.

In summary, theism and science generally support each other, though there
are occasionally conflicts.  Everlasting life has raised a puzzle for
me.  Human and/or divine free will are also puzzling.  Whether God is
seen as probable or improbable depends on one's assumptions.  The problem
of evil may be reformulated as the problem of elegance.

Let me close with an aphorism that I coined to summarize my thoughts as a
scientist and as a Christian:\\

\hspace*{1.0in} {\em Science reveals the intelligence of the universe;}
\\ \hspace*{1.0in} {\em the Bible reveals the Intelligence behind the
universe.}

\newpage
\section*{Acknowledgments}

\hspace{.20in}	I am indebted to discussions with Andreas Albrecht, Denis
Alexander, Stephen Barr, John Barrow, Nick Bostrom, Raphael Bousso,
Andrew Briggs, Peter Bussey, Bernard Carr, Sean Carroll, Brandon Carter,
Kelly James Clark, Gerald Cleaver, Francis Collins, Robin Collins, Gary
Colwell, William Lane Craig, Paul Davies, Richard Dawkins, William
Dembski, David Deutsch, the late Bryce DeWitt, Michael Douglas, George
Ellis, Debra Fisher, Charles Don Geilker, Gary Gibbons, J.~Richard Gott,
Thomas Greenlee, Alan Guth, James Hartle, Stephen Hawking, Rodney Holder,
Chris Isham, Werner Israel, Renata Kallosh, Klaas Kraay, Karel
Kucha\v{r}, Denis Lamoureux, John Leslie, Andrei Linde, Robert Mann, Don
Marolf, Alister McGrath, Ernan McMullin, Gerard Nienhuis, Andrew Page,
Cathy Page, John Page, Gary Patterson, Alvin Plantinga, Chris Polachic,
John Polkinghorne, Martin Rees, Hugh Ross, Henry F.~Schaefer III, Paul
Shellard, James Sinclair, Lee Smolin, Mark Srednicki, Mel Stewart,
Jonathan Strand, Leonard Susskind, Richard Swinburne, Max Tegmark, Donald
Turner, Neil Turok, Bill Unruh, Alex Vilenkin, Steven Weinberg, Robert
White, and others whom I don't recall right now, on various aspects of
these general issues, though the opinions expressed herein are my own.  I
particularly thank Klaas Kraay for providing me with many references on
multiverse theodicies.  My scientific research on the multiverse is
supported in part by the Natural Sciences and Research Council of Canada.

\end{document}